\documentstyle[floats,aps]{revtex}
\begin{document}
\renewcommand{\thefootnote}{\fnsymbol{footnote}}
\draft
\title{\large\bf 
  Integrable Kondo impurity 
  in one-dimensional $q$-deformed $t-J$ models}

\author { Xiang-Yu Ge \footnote {E-mail: xg@maths.uq.edu.au},
 Mark D. Gould, Jon Links and
Huan-Qiang Zhou \footnote {E-mail: hqz@maths.uq.edu.au}}

\address{Department of Mathematics, The University of Queensland,
		     Brisbane, Qld 4072, Australia}

\maketitle

\vspace{10pt}

\begin{abstract}
Integrable Kondo impurities in two cases of the one-dimensional $q$-deformed
$t-J$ models are studied by means of the boundary ${\bf Z}_2$-graded quantum 
inverse scattering method.  The boundary $K$ matrices depending on the local
magnetic moments of the impurities 
are presented  as  nontrivial realizations of the reflection equation
algebras in an impurity Hilbert space.
Furthermore, these  models are  solved by using the algebraic Bethe ansatz 
method and the Bethe ansatz equations are obtained. 
\end{abstract}


%************************** Text Begins here ******************************

%  Greek letters

\def\a{\alpha}
\def\b{\beta}
\def\d{\delta}
\def\e{\epsilon}
\def\g{\gamma}
\def\k{\kappa}
\def\l{\lambda}
\def\o{\omega}
\def\t{\theta}
\def\s{\sigma}
\def\D{\Delta}
\def\L{\Lambda}

% Shorthands for \begin{equation} and the like

\def\beq{\begin{equation}}
\def\eeq{\end{equation}}
\def\bea{\begin{eqnarray}}
\def\eea{\end{eqnarray}}
\def\ba{\begin{array}}
\def\ea{\end{array}}
\def\no{\nonumber}
\def\le{\langle}
\def\re{\rangle}
\def\lt{\left}
\def\rt{\right}

\newcommand{\sect}[1]{\setcounter{equation}{0}\section{#1}}
\renewcommand{\theequation}{\thesection.\arabic{equation}}
\newcommand{\reff}[1]{eq.~(\ref{#1})}

%\newpage
\vskip.3in

\sect{Introduction\label{int}}

The Kondo problem describing the effect due to the exchange interaction
between magnetic impurities and the conduction electrons plays a
very important role in condensed matter physics \cite {Kon64}. Wilson
\cite {Wil75} developed
a very powerful numerical renormalization group approach, and the
model was also solved by the coordinate Bethe ansatz method \cite 
{Wie83,And83} which
gives the specific heat and magnetization. More recently, a conformal
field theory approach was developed by Affleck and Ludwig \cite {AL91}
based on 
a work by Nozi{\`e}res \cite {Noz74}. In the conventional Kondo problem,
the
interaction between conduction electrons is discarded, due to the
fact that the interacting electron system can be described by a Fermi
liquid. Recently there has been substantial research  devoted to the
investigation
of the theory of impurities coupled to  Luttinger liquids. 
Such a problem was first considered by Lee and Toner \cite {LT92}. 
In order 
to get a full
picture about the critical behaviour
of Kondo impurities coupled to Luttinger liquids, some simple
integrable models
which allow exact solutions are desirable.

Several integrable impurity problems in Luttinger
liquids  describing impurities embedded
in systems of correlated electrons have so far appeared in the
literature. 
Among them are  versions of the supersymmetric $t-J$
model with impurities \cite {SZ55,ZS,LF98,AR,BEF,FLP}. 
Such an idea to incorporate an impurity into a
closed chain dates back to Andrei and Johanesson \cite {AJ84}, and Lee
and Schlottmann \cite{Lee} (see also \cite {ZJ89}). 
However, the models
thus constructed suffer from the lack of backward scattering and result in
a very complicated Hamiltonian which is difficult to justify on physical
grounds. Therefore, as observed by Kane and
Fisher \cite {KF92}, it is advantageous to adopt open boundary
conditions with
the impurities situated at the ends of the chain 
when studying  Kondo impurities coupled to
integrable strongly correlated electron systems \cite{PW97}.

%Furthermore, the Bethe ansatz solution and
%conformal properties of the $q$-deformed $t-J$ model without impurity
%were evaluated by Bariev, Kl\"umper, Schadschneider and Zittartz \cite{bksz}.

In our earlier work \cite{ZG,ZGLG}, we were able to derive in an algebraic
fashion integrable boundary Kondo impurities for the isotropic
supersymmetric $t-J$ model.
In this paper, integrable Kondo impurities with spin-$\frac{1}{2}$ coupled
to the one-dimensional $q$-deformed $t-J$ open chain are constructed 
following our earlier formalism. Our 
new input is to search for integrable boundary $K$ matrices
depending on the local magnetic moments of impurities, 
which arise as a nontrivial realization of the ${\bf Z}_2$-graded 
reflection equation (RE) algebras in a finite dimensional quantum
space,which may be interpreted as an impurity Hilbert space.
It should be emphasized that our new non-c-number boundary 
$K$ matrices are highly nontrivial, 
in the sense that they can not be factorized into the product of a
c-number boundary $K$ matrix and the corresponding local monodromy
matrices. The models we present are  solved by means of the 
algebraic Bethe ansatz method and
the Bethe ansatz equations are derived.

Recently, the work of Frahm and Slavnov \cite{FS} has provided a
representation theoretic explanation for the existence of these
non-regular solutions of the reflection equations. These solutions arise
as a projection of a regular solution (subject to some consistency
requirements) onto a subspace of the associated impurity Hilbert space.
The projection method has the effect of reducing the local symmetry of the
projected boundary operator to some subalgebra of the symmetry algebra
of the original regular solution. This result is entirely consistent
with our findings here for the existence of integrable boundary Kondo
impurities in $q$-deformed $t-J$ models.

Before going further, let us make some comments about the relationship
between our construction and others that have appeared in the literature
\cite{SZ55,ZS,Z79,Z}. In these works,
integrable Kondo-like magnetic impurities were studied in the closed $t-J$ and 
Hubbard chains.
Unfortunately, the arguments in these papers does not appear to be 
mathematically sound.
Here, we will address in particular the algebraic approach adopted in
\cite{Z79,Z} for the case of the $t-J$ model. 
In these cases, the author appealed to the Quantum Inverse Scattering
Method (QISM), 
claiming that the Hamiltonians are derivable from the transfer matrix, without
presenting the impurity monodromy matrix. However, in our opinion,
it is reasonable to question the existence of such an impurity
monodromy matrix in view of the form of the Bethe ansatz solutions
obtained. 
A standard calculation shows
that the presence of the impurity changes the pseudovacuum
eigenvalues, but does not affect the fundamantal commutation relations
of the underlying Yangian algebra.  
In \cite{Z79} and in particular Appendix A of \cite{Z} it is claimed
that the operators $\hat{A}_{12},\, \hat{A}_{21}$ both vanish on the
pseudovacuum in the solution of the $t-J$ model with impurities in the
FFB grading. This implies that the pseudovacuum provides a one
dimensional representation of the $su(2)$ 
sub-Yangian (let us call it $\cal{A}$) 
generated by 
$\hat{A}_{11},\, \hat{A}_{22},\ \hat{A}_{12},\, \hat{A}_{21}$, which in
turn implies that the operators $\hat{A}_{11},\, \hat{A}_{22}$ must take
the same eigenvalue on the pseudovacuum. It is clear from the transfer
matrix eigenvalues presented in \cite{Z79,Z} that this does not occur in
those works. Furthermore, it means that the effect of the impurity only
changes the first level Bethe ansatz equations leaving the second level
nested equations unchanged. Inspection of the Bethe ansatz equations
given in \cite{Z79,Z} again shows an inconsistency since in those papers
it is clear that the impurity affects the Bethe ansatz equations at both
levels. 

On the other hand, if the impurity in the 
monodromy matrix is coming from a higher
dimensional atypical representation of $gl(2|1)$ with a shift in the
spectral parameter, as seems to be
the case in \cite{Z79,Z}, then the operator $\hat{A}_{21}$ {\it does not}
vanish on the pseudovacuum chosen there. 
In this instance, it is necessary to use for the Bethe ansatz procedure
a subspace of pseudovacuum states which are stable under the action of the
$su(2)$ sub-Yangian  
$\cal{A}$, as was adopted in 
\cite{LF98,AR,FLP} for other impurity $t-J$ models. 
A $2l+1$-dimensional atypical representation of
$gl(2|1)$ decomposes into an $l+1$-dimensional (spin $S=l/2$) and an
$l$-dimensional (spin $S=(l-1)/2$) representation with respect to the
$su(2)$ subalgebra.
Only in the case of the fundamental representation ($l=1$) is there a
singlet state with respect to $su(2)$ on which $\hat{A}_{21}$ will
vanish. Performing the Bethe ansatz with these considerations does not
reproduce the transfer matrix eigenvalues and Bethe ansatz equations
given in \cite{Z79,Z}.

As shown in our work \cite{ZG,ZGLG,ZGG1,ZGG2}, for the $t-J$ model $R$
matrix and Shastry's $R$ matrix for the Hubbard model, there are no such local
impurity monodromy matrices to guanrantee that their electron-impurity
scattering matrices can be inferred from the corresponding Hamiltonians
\cite{ZGLG}.
Specifically, although for the $t-J$ model there is a singular local
monodromy matrix, no such local monodromy matrix exists for the Hubbard
model.
Also, even for the $t-J$ model, the singularity of such a local monodromy
matrix
does not allow us to use it to construct a closed $t-J$ chain interacting
with magnetic impurities in a closed chain. This conclusion was
confirmed in \cite{FS}. 

The above argument also clearly indicates that our construction is completely
different from that of \cite{Z79,Z}, because there it was  claimed that the
impurity position in a open chain is immaterial, that is, the Kondo
impurity may be put either in the bulk or at boundaries \cite{Z79}.
Moreover, it was  claimed the impurity is a forward scatterer \cite{ZJ},
which is in contrast to our results.
As shown in \cite{ZG,ZGLG},
for the $t-J$ 
model, such a singular local monodromy
matrix can be used to well-define a non-c-number boundary K matrix
which leads us to integrable Kondo-like impurities in the corresponding
open chains. 
This shows the
important result that this Kondo impurity is  completely backward-scattering.

The layout of this paper is the following.  We begin by reviewing the
${\bf Z}_2$-graded boundary QISM  as
formulated in \cite{Zhou97,BRA97}. We then introduce two integrable
cases of the one-dimensional $q$-deformed $t-J$ model with Kondo impurities
on the
boundaries. Integrability of the models is established by relating the
Hamiltonians to one parameter families of commuting transfer matrices.
This is achieved through solving the reflection equations for 
non-c-number solutions. 
Finally we solve the models by means of the algebraic 
Bethe ansatz method and derive the Bethe ansatz equations.

\sect{Graded Reflection Equation Algebra and Transfer Matrix 
      \label{for}}

In this section, we give a brief review about
the ${\bf Z}_2$-graded boundary quantum inverse scattering method.
To begin, let
$V$ be a finite-dimensional ${\bf Z}_2$-graded linear superspace
and let the operator $R(u)$ satisfy  the graded quantum
Yang-Baxter equation
\bea
R_{12}(u_1-u_2)R_{13}(u_1)R_{23}(u_2)=R_{23}(u_2)
R_{13}(u_1)R_{12}(u_1-u_2).\no
\eea
Here $R_{jk}(u)$ denotes the matrix on $V\otimes V\otimes V$
acting on the
$j$-th and $k$-th superspaces and as an identity on the remaining
superspace. The standard notation is used 
with $R_{12}(u)=R(u) \otimes I, R_{23}(u)=I \otimes R(u)$ and etc.,
where
$R(u)=\sum_i a_i \otimes b_i \in End( V \otimes V)$. The
variables $u_1$ and $u_2$ are spectral parameters.
The tensor product
should be understood in the graded sense, that is the
multiplication rule for any homogeneous elements $x,\,y,\,x',\,y'\in End
V$ is given by
\beq
(x\otimes y) (x'\otimes y')=(-1)^{[y][x']}\;(xx'\otimes yy')
\label{rule} \eeq
where $[x]$ stands for the ${\bf Z}_2$-grading of the element $x$. Let $P$
be the ${\bf Z}_2$-graded permutation operator in $V\otimes V$. Then
$P(x\otimes y)=(-1)^{[x][y]}y\otimes x,~\forall x,y\in V$ and
$R_{21}(u)=P_{12}R_{12}(u)P_{12}$. 

We form the monodromy matrix
$T(u)$ for a $L$-site lattice  chain by  
\bea
T(u)=R_{0L}(u)\cdots R_{01}(u),\no   
\eea
\bea
(T(u)^{ab})_{\a_1,\b_1,\cdots,\a_L,\b_L}&=&R_{0L}(u)^{ac_L}_{\a_L\b_L}
R_{0L-1}(u)^{c_Lc_{L-1}}_{\a_{L-1}\b_{L-1}}\cdots
R_{01}(u)^{c_2b}_{\a_1\b_1}\no\\
&&\times(-1)^{\sum^L_{j=2}([\a_j]+[\b_j])\sum^{j-1}_{i=1}[\a_i]},\no
%\label{3gradL}
\eea
where $0$ still represents the auxiliary superspace, and the
tensor product is still in the graded sense.
$T(u)$ is a quantum operator valued
matrix that acts nontrivially in the graded tensor
product of all quantum superspaces of the lattice.

Indeed, one may
show that $T(u)\in End(V \otimes W), R(u) \in End (V \otimes V)$
generates  a representation of the graded 
quantum Yang-Baxter algebra
\beq
R_{12}(u_1-u_2)\stackrel {1} {T}(u_1) \stackrel{2}{T}(u_2)
  =\stackrel{2}{T}(u_2)\stackrel{1}{T}(u_1) R_{12}(u_1-u_2),
  \label{rtt-ttr}
\eeq
where for notational convenience we have 
$$\stackrel {1} {T}(u)=T_{13}(u),~~~\stackrel{2}{T}(u)=T_{23}(u) $$ 
and the subscript 3 now labels the quantum superspace $W=V^{\otimes
L}$. 

In order to describe integrable Kondo impurities in strongly correlated
electronic models with open boundary
conditions, we need to introduce an appropriate ${\bf Z}_2$-graded 
reflection equation algebra. 
We introduce the associative superalgebras ${\cal T}_-$ and ${\cal T}_+$
defined by the R-matrix and the relations
\bea
R_{12}(u_1-u_2)\stackrel {1} {\cal T}_-(u_1) R_{21}(u_1+u_2)
\stackrel {2}{\cal T}_-(u_2)&=& 
\stackrel {2}{\cal T}_-(u_2) R_{12}(u_1+u_2)
\stackrel {1}{\cal T}_-(u_1) R_{21}(u_1-u_2),\label{re-alg1}
\eea
and
\bea
R_{21}^{st_1 \; ist_2}(-u_1+u_2)\stackrel {1}{{\cal T}^{st_1}_+}(u_1) 
{\tilde R}_{12}(-u_1-u_2)
\stackrel {2}{{\cal T}^{ist_2}_+}(u_2)& = &
\stackrel {2}{{\cal T}^{ist_2}_+}(u_2) {\tilde{\tilde R}}_{21}(-u_1-u_2)
\stackrel {1}{{\cal T}^{st_1}_+}(u_1) R_{12}^{st_1\; ist_2}(-u_1+u_2),
\label{re-alg2}    
\eea
where we have defined  new objects ${\tilde R}$ 
and  ${\tilde {\tilde R}}$ 
through the relations
\bea
  {\tilde R} _{12}^{st_2} (-u_1-u_2) R^{st_1}_{21}(u_1+u_2)&=&1,\no\\
{\tilde {\tilde {R}}} _{21}^{ist_1} (-u_1-u_2) R^{ist_2}_{12}(u_1+u_2)&=&1,
\label{cross}
\eea
and $st_i$ stands for the supertransposition taken in the $i$-th space,
whereas $ist_i$  is the inverse operation of $st_i$. 
One of the important steps towards formulating a
correct formalism for the ${\bf Z}_2$-graded case is to introduce
in the equation (\ref{re-alg2})
the inverse operation of the supertransposition.
In any case, the $R$-matrices enjoy the unitarity property,
\beq
 R_{12}(u_1-u_2)R_{21}(-u_1+u_2) = 1.\label{unitarity} 
\eeq

One can obtain a class of realizations of the superalgebras ${\cal T}_+$  and
${\cal T}_-$  by choosing  ${\cal T}_{\pm}(u)$ to be the form
\beq
{\cal T}_-(u) = T_-(u) \tilde {\cal T}_-(u) T^{-1}_-(-u),~~~~~~ 
{\cal T}^{st}_+(u) = T^{st}_+(u) \tilde {\cal T}^{st}_+(u) 
  \lt(T^{-1}_+(-u)\rt)^{st}\label{t-,t+} 
\eeq
with
\bea
T_-(u) = R_{0M}(u) \cdots R_{01}(u),~~~~
T_+(u) = R_{0L}(u) \cdots R_{0,M+1}(u),~~~~ 
\tilde {\cal T}_{\pm}(u) = K_{\pm}(u),\no
\eea
where $M$ is any index between 1 and $L$, and 
$K_{\pm}(u)$, called boundary $K$-matrices, 
are representations of  ${\cal T}_{\pm}$. 
 In the following, without loss of generality, we shall choose
$M=L$ so that ${\cal T}_+(u)\equiv K_+(u)$.

The $K$-matrices $K_\pm(u)$ satisfy the same
relations as ${\cal T}_\pm(u)$ in (\ref{re-alg1}) and (\ref{re-alg2}),
respectively. 
That is the $K$-matrices
obey the following graded reflection equations:
\bea
R_{12}(u_1-u_2)\stackrel {1} {K}_-(u_1) R_{21}(u_1+u_2)
\stackrel {2}{K}_-(u_2)&=& 
\stackrel {2}{K}_-(u_2) R_{12}(u_1+u_2)
\stackrel {1}{K}_-(u_1) R_{21}(u_1-u_2),\label{RE1}
\eea
and
\bea
R_{21}^{st_1 \; ist_2}(-u_1+u_2)\stackrel {1}{K^{st_1}_+}(u_1) 
{\tilde R}_{12}(-u_1-u_2)
\stackrel {2}{K^{ist_2}_+}(u_2)& = &
\stackrel {2}{K^{ist_2}_+}(u_2) {\tilde{\tilde R}}_{21}(-u_1-u_2)
\stackrel {1}{K^{st_1}_+}(u_1) R_{12}^{st_1\; ist_2}(-u_1+u_2).\no\\
\label{RE2}    
\eea
Now we rewrite the relations (\ref{cross}):
\bea
\tilde{\tilde{R}}_{21}^{{ist_1}\;st_2}(-u_1-u_2)&=&(((R_{21}(-u_1-u_2)^{-1}
 )^{{st_2}^{-1}})^{-1})^{st_2},\no\\
   {\tilde R}_{12}^{{ist_1}\;st_2}(-u_1-u_2)&=&(((R_{12}(-u_1-u_2)^{-1}
       )^{st_1})^{-1})^{{ist_1}}.\no
	   \eea
With the unitarity property (\ref{unitarity}), one can show that
the quantity ${\cal T}^{st}_+(u)$ given by (\ref{t-,t+})
satisfies the equation (\ref{re-alg2}) as the following form,
\bea
&&R_{21}^{st_1 {ist_2}}(-u_1+u_2)\stackrel {1}{{\cal T}^{st_1}_+} (u_1)
  \{\;[\;R^{st_1}_{21}(u_1+u_2)\;]^{-1}\;\}^{{ist_2}}
\stackrel {2}{{\cal T}^{{ist_2}}_+}(u_2)\no\\
&&~~~~~~~~=\stackrel {2}{{\cal T}^{{ist_2}}_+}(u_2)
\{\;[\;R^{{ist_2}}_{12}(u_1+u_2)\;]^{-1}\;\}^{st_1}
  \stackrel {1}{{\cal T}^{st_1}_+}(u_1) R_{12}^{st_1 {ist_2}}(-u_1+u_2).
  \label{bgzzT+}
\eea

The $K$-matrix $K_+(u)$ satisfies the same relation as ${\cal T}_+(u)$
in (\ref{bgzzT+}), which fulfills the following graded
reflection equation
\bea
&&R_{21}^{st_1 {ist_2}}(-u_1+u_2)\stackrel {1}{{K}^{st_1}_+} (u_1)
  \{\;[\;R^{st_1}_{21}(u_1+u_2)\;]^{-1}\;\}^{{ist_2}}
      \stackrel {2}{{K}^{{ist_2}}_+}(u_2)\no\\
	  &&~~~~~~~~=\stackrel {2}{{K}^{{ist_2}}_+}(u_2)
	      \{\;[\;R^{{ist_2}}_{12}(u_1+u_2)\;]^{-1}\;\}^{st_1}
		   \stackrel {1}{{K}^{st_1}_+}(u_1) R_{12}^{st_1
			{ist_2}}(-u_1+u_2).
			     \label{bgzzK+}
				  \eea

Following  Sklyanin's approach in \cite{Skl88},
one defines the boundary transfer matrix $\tau(u)$ as
\bea
\tau(u)&=& str_0 ({\cal T}_+(u){\cal T}_-(u))\no\\
%&=& str (K_+(u){\cal T}(u))\no\\
&=&str_0 \lt(K_+(u)T(u)K_-(u)T^{-1}(-u)\rt).
\label{bgzzstr}
\eea
Then it can be shown that \cite{BRA97}
\bea
[\tau(u_1),\tau(u_2)] = 0.\no
\eea

\sect{Integrable non-c-number boundary $K$-matrices and 
      Kondo impurities in the one-dimensional $t-J$ models \label{Boun}}

Let the operators $c_{j,\sigma }$ and $c_{j,\sigma }^{\dagger }$
denote the
annihilation and creation operators of electron with spin $\sigma $
on a lattice site $j$, and we assume the total number of
lattice sites is $L$, $\sigma =\downarrow,\uparrow $ represent
spin down and up, respectively. These operators are canonical
Fermi operators satisfying anticommutation relations
$\{c_{j,\sigma }^{\dagger },c_{j,\tau }\} =\delta _{ij}
\delta _{\sigma \tau }$.
We denote by $n_{j,\sigma }=c_{j,\sigma }^{\dagger }c_{j,\sigma }$
the number operator for the electron on a site $j$ with
spin $\sigma $, and by
$n_j=n_{j,\downarrow }+n_{j,\uparrow}$ the number operator for the
electron on a site $j$.
The Fock vacuum state $|0\re$ satisfies $c_{j,\sigma }|0\re=0$. 

We consider the
following type of Hamiltonians describing two magnetic impurities
coupled to
a one-dimensional $q$-deformed supersymmetric $t-J$ open chain with
$U_q(gl(2|1))$ symmetry.
Make the identifications:
\bea
      &&|1\re=c_{j,\downarrow}^\dagger|0\re\,,~~~
	  |2\re=c_{j,\uparrow}^\dagger|0\re\,,~~~
	  |3\re=|0\re.\label{tjchoice}
\eea
Then
\bea
H&=&-\sum _{j=1}^{N-1}\{ \sum_{\s=\uparrow,\downarrow}
(c^\dagger_{j,\s}c_{j+1,\s}+h. c.)
(1-n_{j,-\s})(1-n_{j+1,-\s})
-{S}_j ^+{S}_{j+1}^--S_j^-S_{j+1}^+\no\\
& &- q^{-1}n_{j,\uparrow}(1-n_{j+1,\downarrow})
- qn_{j+1,\uparrow}(1-n_{j,\downarrow})
-q(n_{j,\downarrow}+n_{j+1,\downarrow})\}
\no\\
%& &+\frac{q^{c_a+2}(q-q^{-1})(q^2-2q^{c_a}+1)}
%{2(q^{c_a}-q^2)(q^{c_a+2}-1)} n_1\no\\
%& &+\frac{q^{c_a+2}(q-q^{-1})^2}{2(q^{c_a}-q^2)(q^{c_a+2}-1)}
& &+J_a\lt((q-q^{-1})n_{1,\downarrow}
-(\s_a^-S_1^- + \s_a^+S_1^+)
+\s^z_ a(q n_{1,\uparrow}
-q^{-1}n_{1,\downarrow}) \rt)+V_a n_1\no\\
& &+J_b\lt((q-q^{-1})n_{N,\downarrow}
-(\s_b^-S_N^- + \s_b^+S_N^+)
+\s^z_ b(q n_{N,\uparrow}
-q^{-1}n_{N,\downarrow}) \rt)+V_b n_N. 
\label {ham1}
\eea
Here $S^+_j,S^-_j$
as usual is the vector spin operator for
 the conduction electrons at site $j$
and expressed as $S^+_j=c^\dagger_{j,\uparrow}c_{j,\downarrow},
S^-_j=c^\dagger_{j,\downarrow}c_{j,\uparrow}$;
 $\s^{\pm}_{g}=\s^x_{g} \pm
  i\s^y_{g}, \s^z_{g} (g = a,b)$ are the local
  moments with spin-$\frac{1}{2}$ located at the left and right ends of
  the system respectively. The Kondo coupling constants $J_g, V_g
  (g=a,b)$ at the left and right ends of the chain are expressed in
  terms
  of the arbitrary parameters $c_g$ in the form
  \bea
  J_g=\frac{q^{c_g+2}(q-q^{-1})^2}{2(q^{c_g}-q^2)(q^{c_g+2}-1)},~~~~
  V_g=\frac{q^{c_g+2}(q-q^{-1})(q^2-2q^{c_g}+1)}
  {2(q^{c_g}-q^2)(q^{c_g+2}-1)}. \no
  \eea

It has been shown in 
ref. \cite {FK93a} that the bulk Hamiltonian acquires an
underlying supersymmetry algebra given by $U_q(gl(2|1))$ in the minimal
representation. Throughout we will refer to this case as the
supersymmetric $t-J$ model. Integrability of this model 
on the open chain
with free boundary conditions was established by 
Foerster and Karowski
\cite {FK93a} by showing that the model can be constructed using the
QISM. Furthermore, open chain integrability with appropriate
boundary conditions was shown in refs. \cite{Gon94a}.  

It is quite interesting to note that although
the introduction  of integrable  impurities we propose below 
spoils the supersymmetry, there 
still remains  $U_q(su (3))$ symmetry in the Hamiltonian (\ref {ham2}) which
maintains conservation of total spin and electron number.
We will also establish the quantum integrability of the following Hamiltonian:
\bea
H&=&-\sum _{j=1}^{N-1}\{ \sum_{\s=\uparrow,\downarrow}
(c^\dagger_{j,\s}c_{j+1,\s}+h. c.)
(1-n_{j,-\s})(1-n_{j+1,-\s})
+{S}_j ^+{S}_{j+1}^-+S_j^-S_{j+1}^+\no\\
& &- q^{-1}n_{j,\uparrow}(1-n_{j+1,\downarrow})
-
qn_{j+1,\uparrow}(1-n_{j,\downarrow})-q(n_{j,\downarrow}+n_{j+1,\downarrow})
\no\\
& &+(q+ q^{-1})(n_{j,\downarrow}n_{j+1,\downarrow}
+n_{j,\uparrow}n_{j+1,\uparrow})\}
\no\\
& &+J_a\lt((q-q^{-1})n_{1,\downarrow}
-(\s_a^-S_1^- + \s_a^+S_1^+)
+\s^z_ a(q^{-1} n_{1,\uparrow}
-q n_{1,\downarrow}) \rt)+V_a n_1\no\\
& &+J_b\lt((q-q^{-1})n_{N,\downarrow}
-(\s_b^-S_N^- + \s_b^+S_N^+)
+\s^z_ b(q^{-1} n_{N,\uparrow}
-q n_{N,\downarrow}) \rt)+V_b n_N. 
\label {ham2}
\eea
In this case the dependence of the
Kondo coupling constants $J_g, V_g (g=a,b)$ depending
on free parameters $c_g$ take the
form
\bea
J_g=\frac{q^{c_g+2}(q-q^{-1})^2}{2(q^{c_g}-q^2)(q^{c_g+2}-1)},~~~~
V_g=\frac{q^{c_g}(q-q^{-1})(q^2-2q^{c_g-2}+1)}
{2(q^{c_g}-q^2)(q^{c_g+2}-1)}. \no
\eea

Let us recall that the local Hamiltonian of
the  supersymmetric $q$-deformed $t-J$ model is derived from an $R$-matrix
satisfying the Yang-Baxter equation which has the form \cite{FK93a} 
\beq
R(u)=  
\left ( \begin {array} {ccccccccc}
q^{u+2}-1&0&0&0&0&0&0&0&0\\
0&q(q^u-1)&0&-q^u(1-q^2)&0&0&0&0&0 \\
0&0&q(q^u-1)&0&0&0&q^u(1-q^2)&0&0\\
0&-(1-q^2)&0&q(q^u-1)&0&0&0&0&0\\
0&0&0&0&q^{u+2}-1&0&0&0&0\\
0&0&0&0&0&q(q^u-1)&0&q^u(1-q^2)&0\\
0&0&1-q^2&0&0&0&q(q^u-1)&0&0\\
0&0&0&0&0&1-q^2&0&q(q^u-1)&0\\
0&0&0&0&0&0&0&0&q^u-q^2
\end {array}  \right ),\label{R1}
\eeq
where $u$ is the spectral parameter, and 
we chose to adopt the fermionic, fermionic and bosonic (FFB) grading 
that means $[|1\re]=[|2\re]=1,\,[|3\re]=0$
on the indices labelling the basis vectors. 

We now solve  (\ref{RE1}) and (\ref{bgzzK+}) 
for $K_-(u)$ and $K_+(u)$. For the quantum $R$-matrix (\ref {R1}),
one may 
check that the boundary $K$-matrix $K_-(u)$ given by
\beq
K_-(u)=   \left ( \begin {array}
{ccc}
A_-(u)&B_-(u)&0\\
C_-(u)&D_-(u)&0\\
0&0&1
\end {array} \right ),\label{k1-}
\eeq
with
\bea
A_-(u)&=&\frac {2q^{2u+4}-q^{2u+2}+q^{2u}-2q^{u+c_a+2}-2q^{u-c_a+2}+q^2+1
+(q^2-1)(q^{2u}-1) {\s}^z_a}
{2(q^{-u+c_a+2}-1)(q^{u-c_a+2}-1)},\no\\
B_-(u)&=&\frac {q(q^2-1)(q^{2u}-1) {\s}^-_a}
{(q^{-u+c_a+2}-1)(q^{u-c_a+2}-1)},\no\\
C_-(u)&=&\frac {q(q^2-1)(q^{2u}-1) {\s}^+_a}
{(q^{-u+c_a+2}-1)(q^{u-c_a+2}-1)},\no\\
D_-(u)&=&\frac {q^{2u+4}+q^{2u+2}-2q^{u+c_a+2}-2q^{u-c_a+2}+q^4-q^2+2
-q^2(q^2-1)(q^{2u}-1) {\s}^z_a}
{2(q^{-u+c_a+2}-1)(q^{u-c_a+2}-1)},\no
\eea
satisfies the graded reflection equation (\ref{RE1}) (See Appendix).
%Here ${\bf \s}^{\pm}={\bf \s}^x \pm i{\bf \s}^y$.
Then the boundary $K$-matrix $K_+(u)$ defined by
\beq
K_+(u)=   \left ( \begin {array}
{ccc}
A_+(u)&B_+(u)&0\\
C_+(u)&D_+(u)&0\\
0&0&1 
\end {array} \right ),\label{k1+}
\eeq
with
\bea
A_+(u)&=&-\frac {q^u(q^{2u+c_b+4}+2q^{2u+c_b}-2q^{u+2}-2q^{u+2c_b}+q^{c_b}
-q^{2u+c_b+2} +q^{c_b+2})+q^{u+c_b}(q^{2u+2}-1)(q^2-1) {\s}^z_b}
{2(q^{u+4}-q^{c_b})(q^u-q^{c_b})},\no\\
B_+(u)&=&\frac {q^{u+c_b}(q^{2u+2}-1)(q^2-1){\s}^-_b}
{(q^{u+4}-q^{c_b})(q^u-q^{c_b})},\no\\
C_+(u)&=&\frac {q^{u+c_b}(q^{2u+2}-1)(q^2-1){\s}^+_b}
{(q^{u+4}-q^{c_b})(q^u-q^{c_b})},\no\\
D_+(u)&=&-\frac {q^u(q^{2u+c_b+4}+q^{2u+c_b+2}-2q^{u+4}-2q^{u+2c_b+2}
+2q^{c_b+4} -q^{c_b+2}+q^{c_b})
-q^{u+c_b}(q^{2u+2}-1)(q^2-1) {\s}^z_b}
{2(q^{u+4}-q^{c_b})(q^u-q^{c_b})},\no
\eea
is a solution of the graded reflection equation (\ref{bgzzK+}).

It can be shown that the 
Hamiltonian (\ref{ham1}) is related to the derivative of the
corresponding boundary transfer matrix
$\tau (u)$ with respect to the spectral parameter $u$ at $u=0$ (up to 
an unimportant additive chemical potential term)
\beq
 -H= \sum _{j=1}^{L-1} H_{j,j+1} + \frac {1}{2} \stackrel {1}{K'}_-(0)
+\frac {str_0 (\stackrel {0}{K}_+(0)H_{L0})}{str_0\stackrel{0}{K}_+(0)}
\label{HAM1} 
\eeq
with
\bea
H_{i,j}=\frac{d}{du}P_{i,j}R_{i,j}(u)\mid_{u=0}=P_{i,j}R'_{i,j}(0),\no
\eea
where $P$ is ${\bf Z}_2$-graded permutation operator with the grading
$P[1]=P[2]=1$ and $P[3]=0$.

The second choice of integrable couplings (\ref{ham2}) results from use of an
$R$-matrix obtained by imposing ${\bf Z}_2$-grading to the fundamental
$q$-deformed $U_q(su(3))$ $R$-matrix which reads
\beq
R(u)=  
\left ( \begin {array} {ccccccccc}
-q^{u}+q^2&0&0&0&0&0&0&0&0\\
0&-q(q^u-1)&0&-q^u(1-q^2)&0&0&0&0&0 \\
0&0&q(q^u-1)&0&0&0&q^u(1-q^2)&0&0\\
0&-(1-q^2)&0&-q(q^u-1)&0&0&0&0&0\\
0&0&0&0&-q^{u}+q^2&0&0&0&0\\
0&0&0&0&0&q(q^u-1)&0&q^u(1-q^2)&0\\
0&0&1-q^2&0&0&0&q(q^u-1)&0&0\\
0&0&0&0&0&1-q^2&0&q(q^u-1)&0\\
0&0&0&0&0&0&0&0&q^u-q^2
\end {array}  \right ),\label{R2}
\eeq
where again $u$ is the spectral parameter and 
we adopt the same choice for the ${\bf Z}_2$-grading of the basis states as
before.

We now solve  (\ref{RE1}) and (\ref{bgzzK+})
for $K_-(u)$ and $K_+(u)$. For the quantum $R$-matrix (\ref {R2}),
one may 
check that the matrix $K_-(u)$ given by (\ref{k1-})
with
\bea
A_-(u)&=&\frac {q^{2u+4}-q^{2u+2}+2q^{2u}-2q^{u+c_a+2}-2q^{u-c_a+2}+q^4+q^2
-q^2(q^2-1)(q^{2u}-1) {\s}^z_a}
{2(q^{-u-c_a+2}-1)(q^{u+c_a+2}-1)},\no\\
B_-(u)&=&-\frac {q(q^2-1)(q^{2u}-1) {\s}^-_a}
{(q^{-u-c_a+2}-1)(q^{u+c_a+2}-1)},\no\\
C_-(u)&=&-\frac {q(q^2-1)(q^{2u}-1) {\s}^+_a}
{(q^{-u-c_a+2}-1)(q^{u+c_a+2}-1)},\no\\
D_-(u)&=&\frac {q^{2u+2}+q^{2u}-2q^{u+c_a+2}-2q^{u-c_a+2}+2q^4-q^2+1
+(q^2-1)(q^{2u}-1) {\s}^z_a}
{2(q^{-u-c_a+2}-1)(q^{u+c_a+2}-1)},\no
\eea
satisfies the graded reflection equation (\ref{RE1}).
For this case $K_+(u)$ defined by (\ref{k1+}) with
\bea
A_+(u)&=&\frac {q^u(2q^{2u+c_b+4}+q^{2u+c_b}-2q^{u+4}-2q^{u+2c_b+6}+q^{c_b+8}
-q^{2u+c_b+2} +q^{c_b+6})-q^{u+c_b}(q^{2u}-q^6)(q^2-1) {\s}^z_b}
{2q^2(q^{u+c_b}-1)(q^{u+c_b}-q^4)},\no\\
B_+(u)&=&-\frac {q^{u+c_b}(q^{2u}-q^6)(q^2-1){\s}^-_b}
{q^2(q^{u+c_b}-1)(q^{u+c_b}-q^4)},\no\\
C_+(u)&=&-\frac {q^{u+c_b}(q^{2u}-q^6)(q^2-1){\s}^+_b}
{q^2(q^{u+c_b}-1)(q^{u+c_b}-q^4)},\no\\
D_+(u)&=&\frac {q^u(q^{2u+c_b+2}+q^{2u+c_b}-2q^{u+2}-2q^{u+2c_b+4}+2q^{c_b+4}
+q^{c_b+8}-q^{c_b+6})
+q^{u+c_b}(q^{2u}-q^6)(q^2-1) {\s}^z_b}
{2q^2(q^{u+c_b}-1)(q^{u+c_b}-q^4)},\no
\eea
is a solution of the graded reflection equation (\ref{bgzzK+}).

It can be shown for this case also that the 
Hamiltonian (\ref{ham2}) can be embedded into the boundary transfer matrix
$\tau (u)$ with respect to the spectral parameter $u$ at $u=0$  
by (\ref{HAM1}).

\sect{The Bethe ansatz solutions \label{bethe}}

Having established the quantum integrability of the models, let us
now
diagonalize the Hamiltonian (\ref {ham1})
by means of the algebraic Bethe ansatz method
\cite {Skl88,Gon94}.
We introduce the `doubled' monodromy matrix ${\cal T}(u)$,
\beq
{\cal T}(u)=T(u)K_-(u)\tilde{T}(u) \equiv
 \left ( \begin {array}
{ccc}
{\cal A}_{11}(u)&{\cal A}_{12}(u)&{\cal B}_1(u)\\
{\cal A}_{21}(u)&{\cal A}_{22}(u)&{\cal B}_2(u)\\
{\cal C}_1(u)&{\cal C}_2 (u)& {\cal D}(u)
\end {array} \right ).\label{u1}
\eeq
where 
$\tilde {T}(u)=T^{-1}(-u)$.
Substituting (\ref{u1}) into the equation (\ref {re-alg1}), 
we may draw the following commutation relations,
\bea
{\check {\cal A}}_{bd}(u_1){\cal C}_c(u_2)&=&\frac {(q^{u_1-u_2+2}-1)
(q^{u_1+u_2}-1)}
{(q^{u_1-u_2}-1)(q^{u_1+u_2}-q^2)}r(u_1+u_2-2)^{eb}_{gh}r(u_1-u_2)^{ih}_{cd}
{\cal C}_e(u_2){\check {\cal A}}_{gi}(u_1)\no\\
& &-\frac {q^{u_1+u_2+1}(1-q^2)(q^{2u_1} -1)(q^{2u_2}-1)}{(q^{u_1+u_2}-q^2)
(q^{2u_1}-q^2)(q^{2u_2}-q^2)}r(2u_1-2)^{gb}_{cd}
{\cal C}_g(u_1) {\cal D}(u_2)\no\\
& &+\frac {q^{u_1-u_2}(1-q^2)(q^{2u_1}-1)}{(q^{u_1-u_2}-1)(q^{2u_1}-q^2)}
r(2u_1-2)^{gb}_{id} {\cal C}_g (u_1) {\check {\cal A}}_{ic}(u_2),\label
{cr1}\\
{\cal D}(u_1){\cal C}_b(u_2) 
&=&\frac {(q^{u_1-u_2+2}-1)(q^{u_1+u_2}-1)}
{(q^{u_1-u_2}-1)(q^{u_1+u_2}-q^2)}
{\cal C}_b(u_2){\cal D}(u_1)-\frac {(1-q^2)(q^{2u_2}-1)}{(q^{2u_2}-q^2)
(q^{u_1-u_2}-1)}
{\cal C}_b(u_1){\cal D}(u_2)\no\\
& &  -\frac {1-q^2}{q^{u_1+u_2}-q^2}{\cal C}_d(u_1)
{\check {\cal A}}_{db}(u_2).\label{cr2}
\eea
Here 
\bea
{\cal A} _{bd}(u) = {\check {\cal A}}_{bd}(u) + \frac {q^{2u}(1-q^2)}
{q^{2u}-q^2} \delta _{bd}{\cal D}(u)\label{cr3}
\eea
and the matrix $r(u)$,
which in turn satisfies the quantum Yang-Baxter
equation 
\bea
r_{12}(u_1-u_2)r_{13}(u_1)r_{23}(u_2)=r_{23}(u_2)r_{13}(u_1)r_{12}(u_1-u_2),
\label{rYBE}
\eea
takes the form
\bea
r^{bb}_{bb}(u)=1,~~~~~r^{12}_{12}(u)=-\frac
{1-q^2}{q^{u+2}-1},~~~~~
r^{21}_{21}(u)=-\frac
{q^u(1-q^2)}{q^{u+2}-1},~~~~~
r^{bd}_{db}(u)=\frac {q(q^u-1)}{q^{u+2}-1},~~(b \neq d,b,d = 1,2).\no
\eea
Choosing the Bethe state $|\Omega \rangle $ as
\bea
|\Omega \rangle = {\cal C}_{d_1}(u_1) \cdots {\cal C}_{d_N}(u_N)
|\Psi\rangle F^{d_1\cdots d_N},\no
\eea
with $|\Psi\rangle $ being the pseudovacuum, the indices $d_j$ run
over the values $1,2$, and $F^{d_1\cdots d_N}$ is a function of the
spectral parameters $u_j$, and applying the boundary transfer
matrix $\tau (u)$
to the state $|\Omega\rangle$, we have
$\tau (u) |\Omega \rangle =\Lambda(u) |\Omega \rangle$, with the
eigenvalue
\bea
\Lambda (u)&=& \frac {q^{2u+2}-1}{q^{2u}-q^2}\frac {q^{u+2}-q^{c_b}}
{q^u-q^{c_b}}
\frac {q^{u}-q^{c_b+2}}{q^{u+4}-q^{c_b}}
(-\frac {q^u(q^u-q^2)}{q^{u+2}-1})^L
\prod ^N_{j=1} \frac {q^2(q^{u+u_j}-1)(q^{u-u_j+2}-1)}
{(q^{u+u_j}-q^2)(q^{u-u_j+2}-q^2)}\no\\
& &-\frac {q^2(q^{2u}-1)}{q^{2u}-q^2}
(\frac {q^2(q^u-1)^2}{(q^{u+2}-1)(q^{-u+2}-1)})^L 
\prod ^N_{j=1} \frac {q^2(q^{u+u_j}-1)(q^{u-u_j+2}-1)}
{(q^{u+u_j}-q^2)(q^{u-u_j+2}-q^2)}
\Lambda ^{(1)}(u;\{u_j\})\no
\eea
provided the parameters $\{ u_j\}$ satisfy
\beq
 \frac {q^{2u_j+2}-1}{q^2(q^{2u_j}-1)}\frac {(q^{u_j+2}-q^{c_b})}
 {(q^{u_j}-q^{c_b})}
\frac {(q^{u_j}-q^{c_b+2})}{(q^{u_j+4}-q^{c_b})}
(\frac {q^{u_j}-q^2}{q(q^{u_j}-1)})^{2L}=\Lambda
^{(1)}(u_j;\{u_i\}). \label {bethe1-1}
\eeq
Here $\Lambda ^{(1)}(u;\{u_j\})$ is the eigenvalue of the nested
boundary transfer
matrix $\tau ^{(1)}(u)$
\bea
\tau^{(1)}(u )=str\lt({K_+^{(1)}}( {u })
T^{(1)}({u}, \{  {u }_j\} )
K_-^{(1)}( {u })
{T^{(1)}}^{-1}(- {u }, \{ {u}_j\} )\rt),\label{nesttau}
\eea
with the fermionic and fermionic (FF) grading
that means $[|1\re]=[|2\re]=1$.  
It arises out of the
$r$ matrices from the first term in the right hand side of (\ref {cr1}).
We can prove that the nested boundary $K$-matrices $K_{\pm}^{(1)}(u)$ are
\beq
K^{(1)}_-(u)=
  \left ( \begin {array}
{cc}
A^{(1)}_-(u)&B^{(1)}_-(u)\\
C^{(1)}_-(u)&D^{(1)}_-(u)
\end {array} \right ),\label{k1-0}
\eeq
with
\bea
A^{(1)}_-(u)&=&\frac {2q^{2u+2}-q^{2u}+q^{2u-2}-2q^{u+c_a}-2q^{u-c_a+2}+q^2+1
+(q^2-1)(q^{2u-2}-1) {\s}^z_a}
{2(q^{-u+c_a+2}-1)(q^{u-c_a+2}-1)},\no\\
B^{(1)}_-(u)&=&\frac {q(q^2-1)(q^{2u-2}-1) {\s}^-_a}
{(q^{-u+c_a+2}-1)(q^{u-c_a+2}-1)},\no\\
C^{(1)}_-(u)&=&\frac {q(q^2-1)(q^{2u-2}-1) {\s}^+_a}
{(q^{-u+c_a+2}-1)(q^{u-c_a+2}-1)},\no\\
D^{(1)}_-(u)&=&\frac {q^{2u+2}+q^{2u}-2q^{u+c_a}-2q^{u-c_a+2}+q^4-q^2+2
-q^2(q^2-1)(q^{2u-2}-1) {\s}^z_a}
{2(q^{-u+c_a+2}-1)(q^{u-c_a+2}-1)}\no
\eea
satisfying the nested graded reflection equation
\bea
r_{12}(u_1-u_2 )\stackrel{1}{K_-^{(1)}}(u_1 )r_{21}(u_1+u_2 )
\stackrel{2}{K_-^{(1)}}(u_2 )
=\stackrel{2}{K_-^{(1)}}(u_2 )
r_{12}(u_1+u_2)\stackrel{1}{K_-^{(1)}}(u_2 ) r_{21}(u_1-u_2 )\label{krre-}
\eea
and
\beq
K^{(1)}_+(u)=
  \left ( \begin {array}
{cc}
A^{(1)}_+(u)&B^{(1)}_+(u)\\
C^{(1)}_+(u)&D^{(1)}_+(u)
\end {array} \right ),\label{k1+0}
\eeq
with
\bea
A^{(1)}_+(u)&=&-\frac {q^u(q^{2u+c_b+4}+2q^{2u+c_b}-2q^{u+2}-2q^{u+2c_b}+q^{c_b}
-q^{2u+c_b+2} +q^{c_b+2})+q^{u+c_b}(q^{2u+2}-1)(q^2-1) {\s}^z_b}
{2(q^{u+4}-q^{c_b})(q^u-q^{c_b})},\no\\
B^{(1)}_+(u)&=&\frac {q^{u+c_b}(q^{2u+2}-1)(q^2-1){\s}^-_b}
{(q^{u+4}-q^{c_b})(q^u-q^{c_b})},\no\\
C^{(1)}_+(u)&=&\frac {q^{u+c_b}(q^{2u+2}-1)(q^2-1){\s}^+_b}
{(q^{u+4}-q^{c_b})(q^u-q^{c_b})},\no\\
D^{(1)}_+(u)&=&-\frac {q^u(q^{2u+c_b+4}+q^{2u+c_b+2}-2q^{u+4}-2q^{u+2c_b+2}
+2q^{c_b+4} -q^{c_b+2}+q^{c_b})
-q^{u+c_b}(q^{2u+2}-1)(q^2-1) {\s}^z_b}
{2(q^{u+4}-q^{c_b})(q^u-q^{c_b})}\no
\eea
satisfying
the nested graded reflection equation
\bea
&&r_{21}^{st_1 {ist_2}}(-u_1+u_2)\stackrel {1}{{K_+^{(1)}}^{st_1}} (u_1)
  \{\;[\;r^{st_1}_{21}(u_1+u_2)\;]^{-1}\;\}^{{ist_2}}
    \stackrel {2}{{K_+^{(1)}}^{{ist_2}}}(u_2)\no\\
    &&~~~~~~~~=\stackrel {2}{{K_+^{(1)}}^{{ist_2}}}(u_2)
    \{\;[\;r^{{ist_2}}_{12}(u_1+u_2)\;]^{-1}\;\}^{st_1}
     \stackrel {1}{{K_+^{(1)}}^{st_1}}(u_1) r_{12}^{st_1
     {ist_2}}(-u_1+u_2).
	   \label{krre+}
	   \eea

For the one-dimensional $q$-deformed supersymmetric $U_q(gl(2|1))$ $t-J$
model $R$-matrix (\ref{R1}), choosing the  pseudovacuum~~$ 
|\Psi\rangle
=(0,0,1)^T$, then
\bea
T_{dd}(u)|\Psi\rangle&=&(q(q^u-1))^L|\Psi\rangle,~~~
T_{33}(u)|\Psi\rangle=(q^u-q^2)^L|\Psi\rangle,\no\\
T_{3d}(u)|\Psi\rangle&\neq& 0,~~~
T_{db}(u)|\Psi\rangle=0,~~~
T_{d3}(u)|\Psi\rangle= 0,\no\\
\tilde{T}_{dd}(u)|\Psi\rangle&=&(-\frac{q^{u+1}(q^u-1)}
{(q^u-q^2)(q^{u+2}-1)})^L|\Psi\rangle,~~~
\tilde{T}_{33}(u)|\Psi\rangle=(-\frac{q^u}{q^{u+2}-1})^L|\Psi\rangle,\no\\
\tilde{T}_{3d}(u)|\Psi\rangle&\neq& 0,~~~
\tilde{T}_{db}(u)|\Psi\rangle=0,~~~
\tilde{T}_{d3}(u)|\Psi\rangle=0,\label{dd33}
\eea
where $d\neq b,~~~ d,b=1,2.$
We also find 
\bea
T_{\a3}(u)\tilde{T}_{3\a}(u)|\Psi\rangle&=&\frac{q^{2u}(1-q^2)}{q^{2u}-q^2}
[\tilde{T}_{33}(u)T_{33}(u)-T_{\a\a} \tilde{T}_{\a\a}(u)]
|\Psi\rangle,\no\\
T_{\a3}(u)\tilde{T}_{3\b}(u)|\Psi\rangle&=&0,~~~~~~\a\neq\b.\label{a33b}
\eea
This leads to  
\bea
{\cal D}(u)|\Psi\rangle&=&
T_{3\a}(u)K_-(u)_{\a\b}\tilde{T}_{\b3}(u)|\Psi\rangle=
(-\frac{q^u(q^u-q^2)}{q^{u+2}-1})^L|\Psi\rangle,\no\\
{\cal B}_d(u)|\Psi\rangle&=&
T_{d\a}(u)K_-(u)_{\a\b}\tilde{T}_{\b3}(u)|\Psi\rangle=0,\no\\
{\cal C}_d(u)|\Psi\rangle&\neq& 0,\no\\
\check {\cal A}_{db}(u)|\Psi\rangle&=&
T_{d\a}(u)K_-(U)_{\a\b}\tilde{T}_{\b b}(u)|\Psi\rangle=
(\frac{q^2(q^u-1)^2}{(q^{u+2}-1)(q^{-u+2}-1)})^L K_-(u)_{db}|\Psi\rangle,\no\\
\check {\cal A}_{dd}(u)|\Psi\rangle&=&
T_{dd}(u)(K_-(u)_{dd}-\frac{q^{2u}(1-q^2)}{q^{2u}-q^2})\tilde{T}_{dd}(u)
|\Psi\rangle
+\frac{q^{2u}(1-q^2)}{q^{2u}-q^2}(-\frac{q^u(q^u-q^2)}{q^{u+2}-1})^L
|\Psi\rangle.\label{dbaa} 
\eea
Here
\bea
{K_-(u)}_{d\a}-\frac{q^{2u}(1-q^2)}{q^{2u}-q^2}=
\frac{q^2(q^{2u}-1)}
{q^{2u}-q^2}{K_-(u-1)}_{d\a},~~~\a=d,b \no
\eea
satisfy the equation (\ref {RE1})
for the reduced problem.
By (\ref{cr3}), we have
\bea
{\check {\cal A}}_{dd}(u)|\Psi\rangle&=&
\frac{q^2(q^{2u}-1)}
{q^{2u}-q^2}{K_-^{(1)}(u)}_{dd}=
(K_-(u)_{dd}-
\frac {q^{2u}(1-q^2)}{q^{2u}-q^2})
(\frac{q^2(q^u-1)^2}{(q^{u+2}-1)(q^{-u+2}-1)})^L |\Psi\rangle,\no\\
{\check {\cal A}}_{db}(u)|\Psi\rangle&=&
\frac{q^2(q^{2u}-1)}
{q^{2u}-q^2}{K_-^{(1)}(u)}_{dd}=
K_-(u)_{db} 
(\frac{q^2(q^u-1)^2}{(q^{u+2}-1)(q^{-u+2}-1)})^L |\Psi\rangle.\label{qaa}
\eea

In our calculation, use of the following relations has also been made:
\bea
q^{2u}(1-q^2)T_{11}(u)\tilde{T}_{11}(u)+q^{2u}(1-q^2)T_{12}(u)\tilde{T}_{21}(u)
+(q^{2u}-q^2)T_{13}(u)\tilde{T}_{31}(u)~~~~~~~~~~~~~~~~~~~~~~~~~~~~~~\no\\
~~~~~~~~~~~~~~~~~~~~~=-(q^{2u+2}-1)\tilde{T}_{31}(u)T_{13}(u)
+q^{2u}(1-q^2)\tilde{T}_{32}(u)T_{23}(u)
+q^{2u}(1-q^2)\tilde{T}_{33}(u)T_{33}(u),\no\\
q^{2u}(1-q^2)T_{11}(u)\tilde{T}_{12}(u)+q^{2u}(1-q^2)T_{12}(u)\tilde{T}_{22}(u)
+(q^{2u}-q^2)T_{13}(u)\tilde{T}_{32}(u)
=-q(q^{2u}-1)\tilde{T}_{32}(u)T_{13}(u),\no\\
q^{2u}(1-q^2)T_{21}(u)\tilde{T}_{11}(u)+q^{2u}(1-q^2)T_{22}(u)\tilde{T}_{21}(u)
+(q^{2u}-q^2)T_{23}(u)\tilde{T}_{31}(u)
=-q(q^{2u}-1)\tilde{T}_{31}(u)T_{23}(u),\no\\
q^{2u}(1-q^2)T_{21}(u)\tilde{T}_{12}(u)+q^{2u}(1-q^2)T_{22}(u)\tilde{T}_{22}(u)
+(q^{2u}-q^2)T_{23}(u)\tilde{T}_{32}(u)~~~~~~~~~~~~~~~~~~~~~~~~~~~~~~\no\\
~~~~~~~~~~~~~~~~~~~~~=(1-q^2)\tilde{T}_{31}(u)T_{13}(u)
-(q^{2u+2}-1)\tilde{T}_{32}(u)T_{23}(u)
+q^{2u}(1-q^2)\tilde{T}_{33}(u)T_{33}(u)\no
\eea
which come from a variant of the (graded) Yang-Baxter algebra 
\beq
\stackrel {1}{T}(u)R(2u)\stackrel {2}{\tilde{T}}(u)=
\stackrel {2}{\tilde{T}}(u)R(2u)\stackrel {1}{T}(u).\label{GYBA}
\eeq

Implementing the change $u \rightarrow u+1$ with respect to the
original problem,
one may check that the nested 
boundary $K$ matrices (\ref{k1-0}) and (\ref{k1+0})
still satisfy the reflection equations (\ref{krre-}) and (\ref{krre+})
for the reduced problem.
After some algebra, the nested boundary
transfer matrix
$ \tau ^{(1)}(u)$ may be recognized as that for the ($N+2$)-site 
$XXZ$~~spin-$\frac {1}{2}$ open chain,
which may be diagonalized following Ref.\cite {Skl88}.
Here we merely give the final result,
\bea
\Lambda ^{(1)}(u;\{ u_j \}) &=&
\frac {(q^{u+2}-q^{c_b})}{(q^u-q^{c_b})}
\frac {(q^{u}-q^{c_b+2})}{(q^{u+4}-q^{c_b})}
\prod _{\a =a,b} \frac {q^{u+c_\a+1}-1}{q^{c_\a}(q^{u-c_\a-1}-q^2)}\no\\
& &\{ \frac {q^{2u+2}-1}{q^2(q^{2u}-1)} \prod _{k=1}^M \frac 
{(q^{u-u_k^{(1)}}-q^2)(q^{u+u_k^{(1)}}-q^2)}
{q^2(q^{u-u_k^{(1)}}-1)(q^{u+u_k^{(1)}}-1)}
+\frac {q^{2u}-q^2}{q^2(q^{2u}-1)}\no\\
& & \prod _{\a =a,b} \frac {q^2(q^{u+c_\a-1}-1 )}{(q^{u+c_\a+1}-1)}
\frac {(q^{u-c_\a-1}-1)}{(q^{u-c_\a+1}-1)}\no\\
& &\prod _{j=1}^N \frac {(q^{u+u_j}-q^2)(q^{u-u_j+2}-q^2)}
{q^2(q^{u+u_j}-1)(q^{u-u_j+2}-1)}
\prod ^{M}_{k=1} \frac {q^2(q^{u-u^{(1)}_k+2}-1)(q^{u+u^{(1)}_k+2}-1)}
{(q^{u-u_k^{(1)}+2}-q^2)(q^{u+u_k^{(1)}+2}-q^2)}\},
\eea
provided the parameters $\{ u_k^{(1)} \}$ satisfy 
\bea
&&\prod _{\a=a,b}
\frac {(q^{u^{(1)}_k+c_\a+1}-q^2)(q^{u^{(1)}_k-c_\a+1}-q^2)}
{q^2(q^{u^{(1)}_k+c_\a+1}-1)(q^{u^{(1)}_k-c_\a+1}-1)}
\prod ^N_{j=1} \frac {(q^{u^{(1)}_k+u_j}-q^2)(q^{u^{(1)}_k-u_j+2}-q^2)}
{q^2(q^{u^{(1)}_k+u_j}-1)(q^{u^{(1)}_k-u_j+2}-1)}\no\\
&&~~~~~~~~~~~ =\prod ^{M}_{\stackrel {l=1}{l \neq k}}
\frac {(q^{u^{(1)}_k-u^{(1)}_l+2}-q^4)(q^{u^{(1)}_k+u^{(1)}_l+2}-q^4)}
 {q^4(q^{u^{(1)}_k-u^{(1)}_l+2}-1)(q^{u^{(1)}_k+u^{(1)}_l+2}-1)}.
\label {bethe1-2}
\eea
After a shift of the parameters $u_j \rightarrow u_j+1,
u_k^{(1)} \rightarrow u^{(1)}_k $, the Bethe ansatz equations (\ref
{bethe1-1}) and 
(\ref {bethe1-2}) may be rewritten as follows
\bea
&&(\frac {q(q^{u_j-1}-1)}{q^{u_j+1}-1})^{2L}
\prod_{\a =a,b}
\frac{q^{c_\a+2}(q^{u_j-c_\a-2}-1)} {q^{u_j+c_\a+2}-1}
=\prod_{k=1}^M \frac{q^2(q^{u_j-u^{(1)}_k-1}-1)}{(q^{u_j-u^{(1)}_k+1}-1)}
\frac{(q^{u_j+u^{(1)}_k-1}-1)}{(q^{u_j+u^{(1)}_k+1}-1)},\no\\
&&\prod_{\a =a,b}\frac{q^2(q^{u^{(1)}_k -c_\a-1}-1)(q^{u^{(1)}_k +c_\a-1}-1)}
{(q^{u^{(1)}_k -c_\a+1}-1)(q^{u^{(1)}_k +c_\a+1}-1)}
\prod_{j=1}^N \frac{q^2(q^{u^{(1)}_k - u_j -1}-1)(q^{u^{(1)}_k+u_j -1}-1)}
{(q^{u^{(1)}_k - u_j +1}-1)(q^{u^{(1)}_k+u_j +1}-1)}\no\\
&&~~~~~~~~~~~~~~ =\prod^M _{\stackrel {l=1}{l \neq k}}
   \frac {q^4(q^{u^{(1)}_k-u^{(1)}_l-2}-1)(q^{u^{(1)}_k+u^{(1)}_l-2}-1)}
   {(q^{u^{(1)}_k-u^{(1)}_l+2}-1)(q^{u^{(1)}_k+u^{(1)}_l+2}-1)}.
   \label{Bethe1}
\eea
or
\def\sh{\sinh}
\bea
&&(\frac {\sinh \gamma(u_j-1)}{\sinh \gamma(u_j+1)})^{2L}
\prod_{\a =a,b}
\frac{\sinh \gamma(u_j-c_\a-2)}{\sh \gamma(u_j+c_\a+2)}
=\prod_{k=1}^M \frac{\sh \gamma(u_j-u^{(1)}_k-1) \sh \gamma(u_j+u^{(1)}_k-1)}
{\sh \gamma(u_j-u^{(1)}_k+1) \sh \gamma(u_j+u^{(1)}_k+1)},\no\\
&&\prod_{\a =a,b}\frac{\sh \gamma(u^{(1)}_k -c_\a-1) 
\sh \gamma(u^{(1)}_k +c_\a-1)}
{\sh \gamma(u^{(1)}_k -c_\a+1) \sh \gamma(u^{(1)}_k +c_\a+1)}
\prod_{j=1}^N \frac{\sh \gamma(u^{(1)}_k - u_j -1) 
\sh \gamma(u^{(1)}_k+u_j -1)}
{\sh \gamma(u^{(1)}_k - u_j +1) \sh \gamma(u^{(1)}_k+u_j +1)}
\no \\
&&~~~~~~~~~~~~~=\prod^M _{\stackrel {l=1}{l \neq k}}
   \frac {\sh \gamma(u^{(1)}_k-u^{(1)}_l-2) 
   \sh \gamma(u^{(1)}_k+u^{(1)}_l-2)}
   {\sh \gamma(u^{(1)}_k-u^{(1)}_l+2) 
   \sh \gamma(u^{(1)}_k+u^{(1)}_l+2)},\no
\eea
with $q=\exp 2\gamma$. The corresponding energy eigenvalue $E$ of the model 
is 
\beq
E=-\sum ^N_{j=1} \frac {4}{\sh \g(u_j-1)\sh \g(u_j+1)}.\label{E}
\eeq
(modulo an unimportant additive constant, which we drop).

We now perform the algebraic Bethe ansatz procedure for the couplings
(\ref{ham2}). Again we  introduce the `doubled' monodromy matrix 
${\cal T}(u)$ as (\ref{u1}).
Substituting (\ref{u1}) into the equation (\ref {re-alg1}), 
we now find the following commutation relations
\bea
{\check {\cal A}}_{bd}(u_1){\cal C}_c(u_2)&=&\frac {(q^{u_1-u_2}-q^2)
(q^{u_1+u_2}-q^4)}
{q^2(q^{u_1-u_2}-1)(q^{u_1+u_2}-q^2)}r(u_1+u_2-2)^{eb}_{gh}r(u_1-u_2)^{ih}_{cd}
{\cal C}_e(u_2){\check {\cal A}}_{gi}(u_1)\no\\
& &+\frac {q^{u_1+u_2}(1-q^2)(q^{2u_1} -q^4)(q^{2u_2}-1)}{(q^{u_1+u_2}-q^2)
(q^{2u_1}-q^2)(q^{2u_2}-q^2)}r(2u_1-2)^{gb}_{cd}
{\cal C}_g(u_1) {\cal D}(u_2)  \no\\
& & -\frac {q^{u_1-u_2}(1-q^2)(q^{2u_1}-q^4)}{q^2(q^{u_1-u_2}-1)(q^{2u_1}-q^2)}
r(2u_1-2)^{gb}_{id} {\cal C}_g (u_1) {\check {\cal A}}_{ic}(u_2),\label
{cr4}
\eea
as well as (\ref{cr2}) and (\ref{cr3}). 
The matrix $r(u)$, 
which also satisfies the quantum Yang-Baxter
equation (\ref{rYBE}), takes the form,
\bea
r^{bb}_{bb}(u)=1,~~~~~r^{12}_{12}(u)=\frac
{1-q^2}{q^{u}-q^2},~~~~~
r^{21}_{21}(u)=\frac
{q^u(1-q^2)}{q^{u}-q^2},~~~~~
r^{bd}_{db}(u)=\frac {q(q^u-1)}{q^{u}-q^2},(b \neq d,b,d = 1,2).\no
\eea
Acting the $\tau(u)$ on the Bethe state $|\Omega \rangle $, 
$|\Omega \rangle = {\cal C}_{d_1}(u_1) \cdots {\cal C}_{d_N}(u_N)
|\Psi\rangle F^{d_1\cdots d_N}$, we have
$\tau (u) |\Omega \rangle =\Lambda(u) |\Omega \rangle$, with the
eigenvalue
\bea
\Lambda (u)& =&
\frac {q^{2u}-q^4}
{q^{2u}-1}
\frac {q^{u+c_b}-1}{q^{u+c_b-2}-1}
\frac {q^{u+c_b-2}-q^2}{q^{u+c_b-2}-q^4}
(-\frac {q^u(q^u-q^2)}{q^{u+2}-1})^L
\prod ^N_{j=1} \frac {q^2(q^{u+u_j}-1)(q^{u-u_j+2}-1)}
{(q^{u+u_j}-q^2)(q^{u-u_j+2}-q^2)}\no\\
& &-\frac {q^2(q^{2u}-1)}{q^{2u}-q^2}
(\frac {q^2(q^u-1)^2}{(q^{u+2}-1)(q^{-u+2}-1)})^L 
\prod ^N_{j=1} \frac {(q^{u+u_j}-q^4)(q^{u-u_j+2}-q^4)}
{q^2(q^{u+u_j}-q^2)(q^{u-u_j+2}-q^2)}
\Lambda ^{(1)}(u;\{u_j\}),\no
\eea
provided the parameters $\{ u_j\}$ satisfy
\beq
 \frac {q^{2u_j}-q^2}{q^{2u_j}-1}\frac {(q^{u_j+c_b}-1)}
 {(q^{u_j+c_b-2}-1)}
\frac {(q^{u_j+c_b-2}-q^2)}{(q^{u_j+c_b-2}-q^4)}
(\frac {q^{u_j}-q^2}{q(q^{u_j}-1)})^{2L}
\prod ^N_{\stackrel {i=1}{i \neq j}}
\frac {q^4(q^{u_j+u_i}-1)(q^{u_j-u_i+2}-1)}
{(q^{u_j+u_i}-q^4)(q^{u_j-u_i+2}-q^4)}
=-\Lambda ^{(1)}(u_j;\{u_i\}). \label {bethe2-1}
\eeq
Here $\Lambda ^{(1)}(u;\{u_i\})$ is the eigenvalue of the nested
boundary transfer
matrix $\tau ^{(1)}(u)$ (\ref{nesttau}),
which arises out of the
$r$ matrices from the first term in the right hand side of (\ref {cr4}),
we can prove that the nested boundary $K$-matrices $K_-^{(1)}(u)$
(\ref{k1-0}), with
\bea
A^{(1)}_-(u)&=&\frac {q^{2u+2}-q^{2u}+2q^{2u-2}-2q^{u+c_a+2}-2q^{u-c_a}+q^4+q^2
-q^2(q^2-1)(q^{2u-2}-1) {\s}^z_a}
{2(q^{-u-c_a+2}-1)(q^{u+c_a+2}-1)},\no\\
B^{(1)}_-(u)&=&-\frac {q(q^2-1)(q^{2u-2}-1) {\s}^-_a}
{(q^{-u-c_a+2}-1)(q^{u+c_a+2}-1)},\no\\
C^{(1)}_-(u)&=&-\frac {q(q^2-1)(q^{2u-2}-1) {\s}^+_a}
{(q^{-u-c_a+2}-1)(q^{u+c_a+2}-1)},\no\\
D^{(1)}_-(u)&=&\frac {q^{2u}+q^{2u-2}-2q^{u+c_a+2}-2q^{u-c_a}+2q^4-q^2+1
+(q^2-1)(q^{2u-2}-1) {\s}^z_a}
{2(q^{-u-c_a+2}-1)(q^{u+c_a+2}-1)},\no
\eea
satisfies the nested graded reflection
equation (\ref{krre-}), and the nested boundary
$K$-matrix $K^{(1)}_+(u)$ (\ref{k1+0}), with
\bea
A^{(1)}_+(u)&=&\frac {q^u(2q^{2u+c_b+4}+q^{2u+c_b}-2q^{u+4}-2q^{u+2c_b+6}
+q^{c_b+8}
-q^{2u+c_b+2} +q^{c_b+6})-q^{u+c_b}(q^{2u}-q^6)(q^2-1) {\s}^z_b}
{2q^2(q^{u+c_b}-1)(q^{u+c_b}-q^4)},\no\\
B^{(1)}_+(u)&=&-\frac {q^{u+c_b}(q^{2u}-q^6)(q^2-1){\s}^-_b}
{q^2(q^{u+c_b}-1)(q^{u+c_b}-q^4)},\no\\
C^{(1)}_+(u)&=&-\frac {q^{u+c_b}(q^{2u}-q^6)(q^2-1){\s}^+_b}
{q^2(q^{u+c_b}-1)(q^{u+c_b}-q^4)},\no\\
D^{(1)}_+(u)&=&\frac {q^u(q^{2u+c_b+2}+q^{2u+c_b}-2q^{u+2}-2q^{u+2c_b+4}
+2q^{c_b+4}
+q^{c_b+8}-q^{c_b+6})
+q^{u+c_b}(q^{2u}-q^6)(q^2-1) {\s}^z_b}
{2q^2(q^{u+c_b}-1)(q^{u+c_b}-q^4)},\no
\eea
satisfies the nested graded reflection equation (\ref{krre+}).
For the one-dimensional $q$-deformed $U_q(su(3))$ $t-J$ model $R$-matrix
(\ref{R2}) with pseudovacuum~~$ |\Psi\rangle
=(0,0,1)^T$ we have the same relations (\ref{dd33}),(\ref{a33b}) and
(\ref{dbaa}) holding true.  Now
\bea
{K_-(u)}_{d\a}-\frac{q^{2u}(1-q^2)}{q^{2u}-q^2}=
\frac{q^2(q^{2u}-1)}
{q^{2u}-q^2}{K_-(u-1)}_{d\a},~~~\a=d,b\no
\eea
satisfy the graded reflection equation (\ref{RE1})
for the reduced problem.
By (\ref{cr3}), we have (\ref{qaa}).
For this calculation, use of the following relations has also been made:
\bea
q^{2u}(1-q^2)T_{11}(u)\tilde{T}_{11}(u)+q^{2u}(1-q^2)T_{12}(u)\tilde{T}_{21}(u)
+(q^{2u}-q^2)T_{13}(u)\tilde{T}_{31}(u)~~~~~~~~~~~~~~~~~~~~~~~~~~~~~~\no\\
~~~~~~~~~~~~~~~~~~~~~~~~~=(q^{2u}-q^2)\tilde{T}_{31}(u)T_{13}(u)
+q^{2u}(1-q^2)\tilde{T}_{32}(u)T_{23}(u)
+q^{2u}(1-q^2)\tilde{T}_{33}(u)T_{33}(u),\no\\
q^{2u}(1-q^2)T_{11}(u)\tilde{T}_{12}(u)+q^{2u}(1-q^2)T_{12}(u)\tilde{T}_{22}(u)
+(q^{2u}-q^2)T_{13}(u)\tilde{T}_{32}(u)
=q(q^{2u}-1)\tilde{T}_{32}(u)T_{13}(u),\no\\
q^{2u}(1-q^2)T_{21}(u)\tilde{T}_{11}(u)+q^{2u}(1-q^2)T_{22}(u)\tilde{T}_{21}(u)
+(q^{2u}-q^2)T_{23}(u)\tilde{T}_{31}(u)
=q(q^{2u}-1)\tilde{T}_{31}(u)T_{23}(u),\no\\
q^{2u}(1-q^2)T_{21}(u)\tilde{T}_{12}(u)+q^{2u}(1-q^2)T_{22}(u)\tilde{T}_{22}(u)
+(q^{2u}-q^2)T_{23}(u)\tilde{T}_{32}(u)~~~~~~~~~~~~~~~~~~~~~~~~~~~~~~~~~\no\\
~~~~~~~~~~~~~~~~~~~~~~~~=(1-q^2)\tilde{T}_{31}(u)T_{13}(u)
+(q^{2u}-q^2)\tilde{T}_{32}(u)T_{23}(u)
+q^{2u}(1-q^2)\tilde{T}_{33}(u)T_{33}(u)
\eea
which as before come from a variant of the graded Yang-Baxter algebra
(\ref{GYBA}).

Implementing the change $u \rightarrow u+1$ with respect to the
original problem,
one may check that the boundary $K$-matrices (\ref{k1-0}) and
(\ref{k1+0}) still satisfy the reflection equations (\ref{krre-}) and
(\ref{krre+}) for the reduced problem.
After some algebra, the nested boundary transfer matrix
$ \tau ^{(1)}(u)$ may be recognized as that for the ($N+2$)-site
$XXZ$~~spin-$\frac {1}{2}$ open chain,
which may be diagonalized following Ref.\cite {Skl88}.
Again we present only the final result
\bea
\Lambda ^{(1)}(u;\{ u_j \}) &=&-
\frac {(q^{u+c_b}-1)}{(q^{u+c_b-2}-1)}
\frac {(q^{u+c_b-2}-q^2)}{(q^{u+c_b-2}-q^4)}
\prod _{\a =a,b} \frac {q^{c_\a}(q^{u-c_\a-1}-q^2)}{q^{u+c_\a+1}-1}\no\\
& &\{ \frac {q^{2u}-q^2}{q^{2u}-1} \prod _{k=1}^M \frac 
{q^2(q^{u-u^{(1)}_k+2}-1)(q^{u+u^{(1)}_k-2}-1)}
{(q^{u-u^{(1)}_k+2}-q^2)(q^{u+u^{(1)}_k-2}-q^2)}
+\frac {q^{2u+2}-1}{q^{2u}-1}\no\\
& & \prod _{\a =a,b} \frac {q^4(q^{u-c_\a-1}-1)(q^{u+c_\a-1}-1)}
 {(q^{u-c_\a-1}-q^2)(q^{u+c_\a-1}-q^2)}\no\\
& &\prod _{j=1}^N \frac {q^2(q^{u+u_j-2}-1)(q^{u-u_j}-1)}
{(q^{u+u_j-2}-q^2)(q^{u-u_j}-q^2)}
\prod ^{M}_{k=1} \frac {(q^{u-u^{(1)}_k}-q^2)(q^{u+u^{(1)}_k-4}-q^2)}
 {q^2(q^{u-u^{(1)}_k}-1)(q^{u+u^{(1)}_k-4}-1)},\no
\eea
provided the parameters $\{ u^{(1)}_k \}$ satisfy 
\bea
&&\prod _{\a=a,b}
\frac {q^4(q^{u^{(1)}_k-c_\a-1}-1)(q^{u^{(1)}_k+c_\a-1}-1)}
{(q^{u^{(1)}_k-c_\a-1}-q^2)(q^{u^{(1)}_k+c_\a-1}-q^2)}
\prod ^N_{j=1} \frac {q^2(q^{u^{(1)}_k+u_j-2}-1)(q^{u^{(1)}_k-u_j}-1)}
{(q^{u^{(1)}_k+u_j-2}-q^2)(q^{u^{(1)}_k-u_j}-q^2)}\no\\
&&~~~~~~~~~~~
=\prod ^{M}_{\stackrel {l=1}{l \neq k}}
\frac {q^4(q^{u^{(1)}_k-u^{(1)}_l+2}-1)(q^{u^{(1)}_k+u^{(1)}_l-2}-1)}
 {(q^{u^{(1)}_k-u^{(1)}_l+2}-q^4)(q^{u^{(1)}_k+u^{(1)}_l-2}-q^4)}.
\label {bethe2-2}
\eea
After a shift of the parameters $u_j \rightarrow u_j+1,
u^{(1)}_k \rightarrow u^{(1)}_k+2 $, the Bethe ansatz equations (\ref
{bethe2-1}) and 
(\ref {bethe2-2}) may be rewritten as follows
\bea
&&(\frac {q(q^{u_j-1}-1)}{q^{u_j+1}-1})^{2L}
\prod_{\a =a,b}
\frac{q^{u_j+c_\a+2}-1 }{q^{c_\a+2}(q^{u_j-c_\a-2}-1)}
   \prod^N _{\stackrel {i=1}{i \neq j}}
   \frac {(q^{u_j+u_i+2}-1)(q^{u_j-u_i+2}-1)}
   {q^4(q^{u_j+u_i-2}-1)(q^{u_j-u_i-2}-1)}\no\\
&&~~~~~~~~~~~
=  \prod_{k=1}^M \frac{(q^{u_j-u^{(1)}_k+1}-1)(q^{u_j+u^{(1)}_k+1}-1)}
{q^2(q^{u_j-u^{(1)}_k-1}-1)(q^{u_j+u^{(1)}_k-1}-1)},\no\\
&&\prod_{\a =a,b}\frac{(q^{u^{(1)}_k-c_\a+1}-1)(q^{u^{(1)}_k+c_\a+1}-1)}
{(q^{u^{(1)}_k-c_\a-1}-1)(q^{u^{(1)}_k+c_\a-1}-1)}
\prod_{j=1}^N \frac{(q^{u_k^{(1)} - u_j +1}-1)(q^{u^{(1)}_k+u_j +1}-1)}
{q^2(q^{u_k^{(1)} - u_j -1}-1)(q^{u^{(1)}_k+u_j -1}-1)}\no\\
&&~~~~~~~~~~~=\prod^M _{\stackrel {l=1}{l \neq k}}
   \frac {(q^{u^{(1)}_k-u^{(1)}_l+2}-1)(q^{u^{(1)}_k+u^{(1)}_l+2}-1)}
    {q^4(q^{u^{(1)}_k-u^{(1)}_l-2}-1)(q^{u^{(1)}_k+u^{(1)}_l-2}-1)},
    \label{Bethe2}
\eea
or
\bea
&&(\frac {\sh \gamma(u_j-1)}{\sh \gamma(u_j+1)})^{2L}
\prod_{\a =a,b}
\frac{\sh \gamma(u_j+c_\a+2)}{\sh \gamma(u_j-c_\a-2)}
   \prod^N _{\stackrel {i=1}{i \neq j}}
   \frac {\sh \gamma(u_j+u_i+2) \sh \gamma(u_j-u_i+2)}
   {\sh \gamma(u_j+u_i-2) \sh \gamma(u_j-u_i-2)}
\no \\
&&~~~~~~~~~=\prod_{k=1}^M \frac{\sh \gamma(u_j-u^{(1)}_k+1) 
\sh \gamma(u_j+u^{(1)}_k+1)}
{\sh \gamma(u_j-u^{(1)}_k-1) \sh \gamma(u_j+u^{(1)}_k-1)},\no\\
&&\prod_{\a =a,b}\frac{\sh \gamma(u^{(1)}_k -c_\a+1) 
\sh \gamma(u^{(1)}_k +c_\a+1)}
{\sh \gamma(u^{(1)}_k -c_\a-1) 
\sh \gamma(u^{(1)}_k +c_\a-1)}
\prod_{j=1}^N \frac{\sh \gamma(u^{(1)}_k - u_j +1)
\sh \gamma(u^{(1)}_k+u_j +1)}
{\sh \gamma(u^{(1)}_k - u_j -1)
\sh \gamma(u^{(1)}_k+u_j -1)}
 \no \\
&&~~~~~~~~~~=\prod^M _{\stackrel {l=1}{l \neq k}}
\frac {\sh \gamma(u^{(1)}_k-u^{(1)}_l+2) \sh \gamma(u^{(1)}_k+u^{(1)}_l+2)}
 {\sh \gamma(u^{(1)}_k-u^{(1)}_l-2) \sh \gamma(u^{(1)}_k+u^{(1)}_l-2)}.\no
\eea
with $\gamma$ and $E$ as before.

\sect{Conclusion \label{con}}

In this paper, we studied the integrability of
the two cases of one-dimensional $q$-deformed
$t-J$ models
with boundary Kondo impurities. The eigenvalues of the Hamiltonian
in each case are derived from
commuting boundary transfer matrices and the Bethe
ansatz equations are obtained by using the algebraic Bethe ansatz
method. Taking the limit $q \rightarrow 1$ in the Bethe ansatz
equations
(\ref{Bethe1}) and (\ref{Bethe2}),
we recover the Bethe ansatz equations
for the two cases of the one-dimensional $gl(2|1)$ and $su(3)$ ~$t-J$
models with the boundary Kondo impurities described when
$s_\a=\frac{1}{2}$ in \cite{ZGLG}. Nevertheless, it would be
interesting to extend the analysis of Bariev {\it et al.} \cite{BKSZ}
to the present
case, which will allow us to extract some exact results about
physical aspects of the models.

After completion of this work, we noticed a preprint from Fan, Wadati
and Yue \cite{FWY},
in which a boundary Kondo impurities  with arbitrary spin is
solved in the one-dimensional generalized supersymmetric $t-J$ model.
However, they did not present the Hamiltonian explicitly and only
treated the case corresponding to $U_q(gl(2|1))$.

\vskip.3in
%\acknowledgments
This work is supported by the Australian Research Council. 

\appendix

\sect{Derivation of the Non-c-Number Boundary
$K$-Matrices for the One-Dimensional $q$-Deformed $t-J$ Models with
Boundary Kondo Impurities}

In this appendix, we sketch the procedure of solving the
graded reflection equation of (\ref{RE1}).
To describe  
the one-dimensional $q$-deformed supersymmetric $U_q(gl(2|1))$~
$t-J$ model with boundary Kondo impurities, it is reasonable to assume that
\beq
K_-(u) =\left (
\begin {array} {ccc}
\bar{A}(u)  &\bar{B}(u)&0\\
\bar{C}(u)&\bar{D}(u) &0\\
0&0&1
\end {array}  \right ).\label{4tjkAK}
\eeq
Choosing $\bar{A}(u)=F^{-1}(u)A(u),~\bar{B}(u)=F^{-1}(u)B(u)$, 
$\bar{C}(u)=F^{-1}(u)C(u),~\bar{D}(u)=F^{-1}(u)D(u)$, then 
\beq
K_-(u) \propto \left (
\begin {array} {ccc}
{A}(u)  &{B}(u)&0\\
{C}(u)&{D}(u) &0\\
0&0&F(u)
\end {array}  \right ).
\eeq
For the $R$-matrix (\ref{R1}), one may get, from the graded
reflection equation (\ref{RE1}),
33 functional equations, of which 11 are identities. After some
algebraic analysis, together with the $U_q(su(2))$ symmetry,
we may assume that
\bea
A(u)&=&\a(u)+\b(u)\s^z, ~~~B(u)=\g(u)\s^-,\no\\
C(u)&=&\g(u)\s^+,~~~
D(u)=\tilde{\a}(u)-\tilde{\b}(u)\s^z. \label {4tjkAK2}
\eea
There are two equations automatically satisfied, leaving only 20 equations
left to be solved
\bea
& &A(u_1)B(u_2)+B(u_1)D(u_2)=
A(u_2)B(u_1)+B(u_2)D(u_1),\no\\
& &C(u_1)A(u_2)+D(u_1)C(u_2)=
C(u_2)A(u_1)+D(u_2)C(u_1),\no\\
& &(q^{u_-}-1)(A(u_1)B(u_2)+B(u_1)D(u_2))=
(q^{u_+}-1)(B(u_1)F(u_2)-q^{u_-}B(u_2)F(u_1)),\no\\
& &(q^{u_-}-1)(A(u_2)B(u_1)+B(u_2)D(u_1))=
(q^{u_+}-1)(B(u_1)F(u_2)-q^{u_-}B(u_2)F(u_1)),\no\\
& &(q^{u_-}-1)(C(u_1)A(u_2)+D(u_1)C(u_2))=
(q^{u_+}-1)(C(u_1)F(u_2)-q^{u_-}C(u_2)F(u_1)),\no\\
& &(q^{u_-}-1)(C(u_2)A(u_1)+D(u_2)C(u_1))=
(q^{u_+}-1)(C(u_1)F(u_2)-q^{u_-}C(u_2)F(u_1)),\no\\
& &(q^{u_-}-1)(A(u_1)A(u_2)+B(u_1)C(u_2)-q^{u_+}F(u_1)F(u_2))=
(q^{u_+}-1)(A(u_1)F(u_2)-q^{u_-}A(u_2)F(u_1)),\no\\
& &(q^{u_-}-1)(A(u_2)A(u_1)+B(u_2)C(u_1)-q^{u_+}F(u_1)F(u_2))=
(q^{u_+}-1)(A(u_1)F(u_2)-q^{u_-}A(u_2)F(u_1)),\no\\
& &(q^{u_-}-1)(C(u_1)B(u_2)+D(u_1)D(u_2)-q^{u_+}F(u_1)F(u_2))=
(q^{u_+}-1)(D(u_1)F(u_2)-q^{u_-}D(u_2)F(u_1)),\no\\
& &(q^{u_-}-1)(C(u_2)B(u_1)+D(u_2)D(u_1)-q^{u_+}F(u_1)F(u_2))=
(q^{u_+}-1)(D(u_1)F(u_2)-q^{u_-}D(u_2)F(u_1)),\no\\
& &(q^{u_-}-1)\lt((q^{u_++2}-1)B(u_1)D(u_2)-(1-q^2)A(u_1)B(u_2)\rt)\no\\
&&~~~~~~=
(q^{u_+}-1)\lt((q^{u_-+2}-1)D(u_2)B(u_1)+q^{u_-}(1-q^2)D(u_1)B(u_2)\rt),\no\\
& &(q^{u_-}-1)\lt((q^{u_++2}-1)D(u_2)C(u_1)-(1-q^2)C(u_2)A(u_1)\rt)\no\\
&&~~~~~~=
(q^{u_+}-1)\lt((q^{u_-+2}-1)C(u_1)D(u_2)+q^{u_-}(1-q^2)C(u_2)D(u_1)\rt),\no\\
& &(q^{u_-}-1)\lt((q^{u_++2}-1)C(u_1)A(u_2)-q^{u_+}(1-q^2)D(u_1)C(u_2)\rt)\no\\
&&~~~~~~=
(q^{u_+}-1)\lt((q^{u_-+2}-1)A(u_2)C(u_1)+(1-q^2)A(u_1)C(u_2)\rt),\no\\
& &(q^{u_-}-1)\lt((q^{u_++2}-1)A(u_2)B(u_1)-q^{u_+}(1-q^2)B(u_2)D(u_1)\rt)\no\\
&&~~~~~~=
(q^{u_+}-1)\lt((q^{u_-+2}-1)B(u_1)A(u_2)+(1-q^2)B(u_2)A(u_1)\rt),\no\\
& &(q^{u_-}-1)\lt((1-q^2)(A(u_1)A(u_2)-q^{u_+}D(u_2)D(u_1))
-(q^{u_++2}-1)(B(u_1)C(u_2) -C(u_2)B(u_1))\rt)\no\\
&&~~~~~=
(q^{u_+}-1)(1-q^2)(D(u_2)A(u_1)-q^{u_-}D(u_1)A(u_2))\no\\
& &(q^{u_-}-1)\lt((1-q^2)(A(u_1)A(u_2)-q^{u_+}D(u_2)D(u_1))
-(q^{u_++2}-1)(B(u_2)C(u_1) -C(u_1)B(u_2))\rt)\no\\
&&~~~~~=
(q^{u_+}-1)(1-q^2)(D(u_2)A(u_1)-q^{u_-}D(u_1)A(u_2))\no\\
&&(q^{u_++2}-1)\lt((1-q^2)B(u_1)D(u_2)+(q^{u_-+2}-1)B(u_2)D(u_1)\rt)\no\\
&&~~~~~=q^{u_+}(1-q^2)\lt((q^{u_-+2}-1)A(u_2)B(u_1)+(1-q^2)A(u_1)B(u_2)\rt)\no\\
&&~~~~~~~~~+q^2(q^{u_+}-1)(q^{u_-}-1)D(u_1)B(u_2)\no\\
&&(q^{u_++2}-1)\lt(q^{u_-}(1-q^2)C(u_1)A(u_2)+(q^{u_-+2}-1)C(u_2)A(u_1)\rt)\no\\
&&~~~~~=q^{u_+}(1-q^2)\lt((q^{u_-+2}-1)D(u_2)C(u_1)+q^{u_-}(1-q^2)
D(u_1)C(u_2)\rt)
\no\\
&&~~~~~~~~~+q^2(q^{u_+}-1)(q^{u_-}-1)A(u_1)C(u_2)\no\\
&&(q^{u_++2}-1)\lt(q^{u_-}(1-q^2)A(u_2)B(u_1)+(q^{u_-+2}-1)A(u_1)B(u_2)\rt)\no\\
&&~~~~~=q^{u_+}(1-q^2)\lt((q^{u_-+2}-1)B(u_1)D(u_2)+q^{u_-}(1-q^2)B(u_2)
D(u_1)\rt)
\no\\
&&~~~~~~~~~+q^2(q^{u_+}-1)(q^{u_-}-1)B(u_2)A(u_1)\no\\
&&(q^{u_++2}-1)\lt((1-q^2)D(u_2)C(u_1)+(q^{u_-+2}-1)D(u_1)C(u_2)\rt)\no\\
&&~~~~~=q^{u_+}(1-q^2)\lt((q^{u_-+2}-1)C(u_1)A(u_2)+(1-q^2)C(u_2)A(u_1)\rt)\no\\
&&~~~~~~~~~+q^2(q^{u_+}-1)(q^{u_-}-1)C(u_2)D(u_1).\no
\eea
%\normalsize
%\noindent
with $u_+=u_1+u_2,u_-=u_1-u_2$.
Substituting (\ref {4tjkAK2}) into these equations,
we find that all these equations are
reduced to the following 11 equations
%{\footnotesize
\bea
&&(\a(u_1)-\b(u_1))\g(u_2)+\g(u_1)(\tilde{\a}(u_2)-\tilde{\b}(u_2))=
(\a(u_2)-\b(u_2))\g(u_1)+\g(u_2)(\tilde{\a}(u_1)-\tilde{\b}(u_1)),\no\\
&&(1-q^2)\lt((1-q^2)(\a(u_1)-\b(u_1))\g(u_2)+(q^{u_-+2}-1)(\a(u_2)-\b(u_2))
\g(u_1)\rt)\no\\
&&~~~~~=(q^{u_++2}-1)\lt((q^{u_-+2}-1)\g(u_2)(\tilde{\a}(u_1)-\tilde{\b}(u_1))
+(1-q^2)\g(u_1)(\tilde{\a}(u_2)-\tilde{\b}(u_2))\rt)\no\\
&&~~~~~~~~~-q^2(q^{u_-}-1)(q^{u_+}-1)(\tilde{\a}(u_1)+\tilde{\b}(u_1))
\g(u_2),\no\\
&&q^{u_+}(1-q^2)\lt(q^{u_-}(1-q^2)\g(u_2)(\tilde{\a}(u_1)-\tilde{\b}(u_1))
+(q^{u_-+2}-1)\g(u_1)(\tilde{\a}(u_2)-\tilde{\b}(u_2))\rt)\no\\
&&~~~~~=(q^{u_++2}-1)\lt((q^{u_-+2}-1)({\a}(u_1)-{\b}(u_1))\g(u_2)
+q^{u_-}(1-q^2)({\a}(u_2)-{\b}(u_2))\g(u_1)\rt)\no\\
&&~~~~~~~~~-q^2(q^{u_-}-1)(q^{u_+}-1)\g(u_2)({\a}(u_1)+{\b}(u_1)),\no\\
&&(q^{u_+}-1)\lt((q^{u_-+2}-1)\g(u_1)(\a(u_2)+\b(u_2))+(1-q^2)\g(u_2)
(\a(u_1)+\b(u_1))
\rt)\no\\
&&~~~~~=(q^{u_-}-1)\lt((q^{u_++2}-1)(\a(u_2)-\b(u_2))\g(u_1)-
q^{u_+}(1-q^2)\g(u_2)(\tilde{\a}(u_1)-\tilde{\b}(u_1))\rt),\no\\
&&(q^{u_+}-1)\lt((q^{u_-+2}-1)(\tilde{\a}(u_2)+\tilde{\b}(u_2))\g(u_1)
+q^{u_-}(1-q^2)(\tilde{\a}(u_1)+\tilde{\b}(u_1))\g(u_2) \rt)\no\\
&&~~~~~=(q^{u_-}-1)\lt((q^{u_++2}-1)\g(u_1)(\tilde{\a}(u_2)-\tilde{\b}(u_2))-
(1-q^2)({\a}(u_1)-{\b}(u_1))\g(u_2)\rt),\no\\
&&(q^{u_+}-1)(1-q^2)\lt((\tilde{\a}(u_2)-\tilde{\b}(u_2))(\a(u_1)+\b(u_1))
-q^{u_-}(\tilde{\a}(u_1)-\tilde{\b}(u_1))(\a(u_2)+\b(u_2))\rt),\no\\
&&~~~~~=(q^{u_-}-1)(1-q^2)\lt((\a(u_1)+\b(u_1))(\a(u_2)+\b(u_2))
-q^{u_+}(\tilde{\a}(u_2)-\tilde{\b}(u_2))(\tilde{\a}(u_1)-\tilde{\b}(u_1))\rt)
\no\\
&&~~~~~~~~+(q^{u_-}-1)(q^{u_++2}-1)\g(u_1)\g(u_2),\no\\
&&(q^{u_+}-1)(1-q^2)\lt((\tilde{\a}(u_2)+\tilde{\b}(u_2))(\a(u_1)-\b(u_1))
-q^{u_-}(\tilde{\a}(u_1)+\tilde{\b}(u_1))(\a(u_2)-\b(u_2))\rt),\no\\
&&~~~~~=(q^{u_-}-1)(1-q^2)\lt((\a(u_1)-\b(u_1))(\a(u_2)-\b(u_2))
-q^{u_+}(\tilde{\a}(u_2)+\tilde{\b}(u_2))(\tilde{\a}(u_1)+\tilde{\b}(u_1))\rt)
\no\\
&&~~~~~~~~-(q^{u_-}-1)(q^{u_++2}-1)\g(u_1)\g(u_2),\no\\
&&(q^{u_-}-1)\lt((\a(u_1)-\b(u_1))(\a(u_2)-\b(u_2))
-q^{u_+}F(u_1)F(u_2)+\g(u_1)\g(u_2)\rt)\no\\
&&~~~~~=(q^{u_+}-1)\lt((\a(u_1)-\b(u_1))F(u_2)
-q^{u_-}(\a(u_2)-\b(u_2))F(u_1)\rt),\no\\
&&(q^{u_-}-1)\lt((\tilde{\a}(u_1)-\tilde{\b}(u_1))(\tilde{\a}(u_2)
-\tilde{\b}(u_2))
-q^{u_+}F(u_1)F(u_2)+\g(u_1)\g(u_2)\rt)\no\\
&&~~~~~=(q^{u_+}-1)\lt((\tilde{\a}(u_1)-\tilde{\b}(u_1))F(u_2)
-q^{u_-}(\tilde{\a}(u_2)-\tilde{\b}(u_2))F(u_1)\rt),\no\\
&&(q^{u_-}-1)\lt((\a(u_1)+\b(u_1))(\a(u_2)+\b(u_2))
-q^{u_+}F(u_1)F(u_2)\rt)\no\\
&&~~~~~=(q^{u_+}-1)\lt((\a(u_1)+\b(u_1))F(u_2)
-q^{u_-}(\a(u_2)+\b(u_2))F(u_1)\rt),\no\\
&&(q^{u_-}-1)\lt((\tilde{\a}(u_1)+\tilde{\b}(u_1))(\tilde{\a}(u_2)
+\tilde{\b}(u_2))
-q^{u_+}F(u_1)F(u_2)\rt)\no\\
&&~~~~~=(q^{u_+}-1)\lt((\tilde{\a}(u_1)+\tilde{\b}(u_1))F(u_2)
-q^{u_-}(\tilde{\a}(u_2)+\tilde{\b}(u_2))F(u_1)\rt).\no
\eea
%\normalsize
%\noindent
Solving these equations using some nontrivial tricks of variable
separation, we have
\bea
\a(u)+\b(u)&=&\tilde{\a}(u)+\tilde{\b}(u)=
\frac{(q^{u+2}-q^{c})(q^{u+2}-q^{-c})}
{(q^{u+c}-q^2)(q^{u-c}-q^2)},\no\\
\a(u)-\b(u)&=&\frac{q^{2u+4}-q^{2u+2}+q^{2u}-q^{u+c+2}-q^{u-c+2}+q^2}
{(q^{u+c}-q^2)(q^{u-c}-q^2)},\no\\
\tilde{\a}(u)-\tilde{\b}(u)&=&
\frac{q^{2u+2}-q^{u+c+2}-q^{u-c+2}+q^4-q^2+1}
{(q^{u+c}-q^2)(q^{u-c}-q^2)},\no\\
\g(u)&=& \frac{q(q^{2}-1)(q^{2u}-1)}
{(q^{u+c}-q^2)(q^{u-c}-q^2)},\no\\
F(u)&=& \frac{(1-\xi q^{-u})} {(1-\xi q^{u})}
\frac{(q^{u+2}-q^{c})(q^{u+2}-q^{-c})}
{(q^{u+c}-q^2)(q^{u-c}-q^2)},\no
\eea
\bea
\xi=q^{c+2},~~~or~~~\xi=q^{-c+2},
\eea
Choosing $\xi=q^{c+2}$, then substituting these 
results into (\ref{4tjkAK2}) and (\ref{4tjkAK}), we may establish
the non-c-number
boundary $K$-matrix $K_-(u)$ (\ref {k1-}).  
Using the same method, we may find other non-c-number boundary $K$-matrix
$K_+(u)$ (\ref{k1+}) by solving the dual graded reflection equation 
(\ref{bgzzK+}).

A similar construction also works for the one-dimensional $q$-deformed 
$U_q(su(3))$ ~$t-J$ model with the boundary Kondo impurities with
the quantum $R$ matrix (\ref{R2}).

%\newpage
%\vskip.3in

\end{document}